# Comparative Analysis of Machine Learning Models for Lung Cancer Mutation Detection and Staging Using 3D CT Scans

Yiheng Li[1], Francisco Carrillo-Perez[1], Mohammed Alawad[2], Olivier Gevaert[1,3]

[1]Stanford Center for Biomedical Informatics Research (BMIR), Department of Medicine, Stanford University, Stanford, CA 94305, USA
[2]National Center for AI (NCAI), Saudi Data and AI Authority (SDAIA), Riyadh, Saudi Arabia
[3]Department of Biomedical Data Science, Stanford University, Stanford, CA 94305, USA
*Correspondence: ogevaert@stanford.edu

## Abstract

Lung cancer is the leading cause of cancer mortality worldwide, and non-invasive methods for detecting key mutations and staging are essential for improving patient outcomes. Here, we compare the performance of two machine learning models—FMCIB+XGBoost, a supervised model with domain-specific pretraining, and Dinov2+ABMIL, a self-supervised model with attention-based multiple-instance learning—on 3D lung nodule data from the Stanford Radiogenomics and Lung-CT-PT-Dx cohorts. In the task of KRAS and EGFR mutation detection, FMCIB+XGBoost consistently outperformed Dinov2+ABMIL, achieving accuracies of 0.846 and 0.883 for KRAS and EGFR mutations, respectively. In cancer staging, Dinov2+ABMIL demonstrated competitive generalization, achieving an accuracy of 0.797 for T-stage prediction in the Lung-CT-PT-Dx cohort, suggesting SSL's adaptability across diverse datasets. Our results emphasize the clinical utility of supervised models in mutation detection and highlight the potential of SSL to improve staging generalization, while identifying areas for enhancement in mutation sensitivity.

# Introduction

Lung cancer remains the leading cause of cancer-related mortality worldwide, responsible for over 1.8 million deaths each year. The primary challenges in managing lung cancer lie in its early detection and accurate staging, both of which are critical to personalizing treatment and enhancing patient survival. Recent advancements in understanding the genetic and molecular landscape of lung cancer have underscored the importance of mutations, such as those in the *KRAS* and *EGFR* genes, which are linked to unique tumor characteristics and may serve as biomarkers to guide targeted therapies. Traditionally, mutation detection and cancer staging rely on invasive biopsy methods, which, despite their diagnostic accuracy, carry inherent risks and limitations for patients. As a result, there is growing interest in developing non-invasive techniques that leverage the power of advanced imaging and machine learning.

The intersection of radiology and genomics, termed radiogenomics, has gained traction as a field that combines imaging and molecular data to reveal unique insights into tumor biology. With the increasing availability of annotated imaging datasets, machine learning (ML) offers a powerful tool to explore these associations and detect complex patterns. However, traditional supervised ML models require substantial amounts of labeled data, limiting their scalability across diverse clinical settings. This constraint has led to a growing interest in self-supervised learning (SSL) approaches, where models can learn from large volumes of unlabeled data, potentially capturing robust, generalized features that reflect underlying biological information.

This study leverages the potential of both traditional supervised and SSL-based approaches in lung cancer radiogenomics. We evaluate two distinct ML frameworks: (1) FMCIB+XGBoost, which utilizes a foundation model - FMCIB - pre-trained on radiology-specific data as a feature extractor, followed by a fine-tuned XGBoost classifier for classification tasks, and (2) Dinov2+ABMIL, an SSL model pre-trained on natural image datasets, paired with an Attention-Based Multiple Instance Learning (ABMIL) classifier. This approach enables us to assess the effectiveness of SSL-based feature extraction for mutation detection and staging, where instance-level labels are often unavailable.

The comparative evaluation of these models across two independent cohorts provides insights into their performance variability and generalization capability. We hypothesize that the domain-specific FMCIB model, trained on radiology data, will consistently outperform general-purpose models in all tasks due to its tailored feature extraction capabilities. However, we aim to investigate the performance of the general-purpose Dinov2 SSL model, pretrained on natural images, to assess its adaptability to domain-specific tasks, thereby highlighting its potential generalizability. Additionally, we compare the efficacy of 3D models against multi-instance learning approaches using 2D slices, focusing on how these architectures handle spatial and contextual information in complex radiology tasks.

By exploring these models' strengths and limitations, this work aims to contribute to the development of clinically relevant, non-invasive biomarkers for lung cancer management, highlighting the promise and challenges of SSL in medical imaging.

# Results

## KRAS Mutation Detection

The FMCIB+XGBoost model outperformed Dinov2+ABMIL in detecting KRAS mutations within the Stanford Radiogenomics cohort across most performance metrics. FMCIB+XGBoost achieved higher accuracy (0.846 ± 0.036 vs. 0.773 ± 0.009), F1-score (0.716 ± 0.079 vs. 0.436 ± 0.002), and AUC (0.689 ± 0.084 vs. 0.498 ± 0.005). Notably, FMCIB+XGBoost demonstrated a sensitivity of 0.406 ± 0.174, while Dinov2+ABMIL showed no sensitivity, indicating FMCIB's ability to identify positive cases more effectively.

Dinov2+ABMIL, however, exhibited a higher specificity (0.996 ± 0.011 vs. 0.973 ± 0.021), reflecting its capability to correctly identify negative cases. While Dinov2+ABMIL maintained strong specificity, its lack of sensitivity suggests that the model had difficulty predicting the positive cases and it limited its overall effectiveness in KRAS mutation detection compared to the more balanced performance of FMCIB+XGBoost.

## EGFR Mutation Detection

For EGFR mutation detection, the FMCIB+XGBoost model demonstrated superior performance across all key metrics compared to Dinov2+ABMIL. FMCIB+XGBoost achieved an AUC of 0.736 ± 0.073, indicating stronger ability to distinguish between mutation-positive and mutation-negative cases, compared to the much lower AUC of 0.498 ± 0.003 for Dinov2+ABMIL which is roughly random guess. In terms of accuracy, FMCIB+XGBoost achieved 0.883 ± 0.033, surpassing Dinov2+ABMIL's 0.798 ± 0.008.

FMCIB+XGBoost also exhibited a higher F1-score (0.774 ± 0.071 vs. 0.444 ± 0.002), reflecting its better balance of precision and recall. Sensitivity was another area where FMCIB+XGBoost excelled, achieving 0.492 ± 0.143, while Dinov2+ABMIL showed no sensitivity (0.0 ± 0.0). Both models maintained high specificity, with Dinov2+ABMIL slightly outperforming at 0.996 ± 0.007 compared to FMCIB+XGBoost's 0.980 ± 0.015.

Overall, these results indicate that FMCIB+XGBoost is more effective in identifying EGFR mutations, benefiting from domain-specific training, whereas Dinov2+ABMIL struggled to capture mutation-specific features and the model has difficulty predicting positive cases.

| Mutation/Stage | Model | Accuracy | F1-Score | Sensitivity | Specificity | AUC |
| --- | --- | --- | --- | --- | --- | --- |
| **KRAS** | **FMCIB+XGBoost** | 0.846 ± 0.036 | 0.716 ± 0.079 | 0.406 ± 0.174 | 0.973 ± 0.021 | **0.689 ± 0.084** |
| | Dinov2+AB | 0.773 ± | 0.436 ± | 0.0 ± 0.0 | 0.996 ± 0.011 | 0.498 ± |

| | | Accuracy | F1-Score | Sensitivity | Specificity | AUC |
|---|---|---|---|---|---|---|
| | MIL | 0.009 | 0.002 | | | 0.005 |
| EGFR | FMCIB+XGBoost | 0.883 ± 0.033 | 0.774 ± 0.071 | 0.492 ± 0.143 | 0.980 ± 0.015 | **0.736 ± 0.073** |
| | Dinov2+ABMIL | 0.798 ± 0.008 | 0.444 ± 0.002 | 0.0 ± 0.0 | 0.996 ± 0.007 | 0.498 ± 0.003 |
| T Stage | FMCIB+XGBoost | 0.854 ± 0.056 | 0.853 ± 0.057 | 0.843 ± 0.069 | 0.867 ± 0.067 | **0.855 ± 0.056** |
| | Dinov2+ABMIL | 0.598 ± 0.061 | 0.588 ± 0.060 | 0.568 ± 0.178 | 0.634 ± 0.148 | 0.601 ± 0.055 |
| N Stage | FMCIB+XGBoost | 0.890 ± 0.042 | 0.808 ± 0.087 | 0.558 ± 0.173 | 0.987 ± 0.016 | **0.772 ± 0.087** |
| | Dinov2+ABMIL | 0.798 ± 0.034 | 0.576 ± 0.104 | 0.186 ± 0.149 | 0.975 ± 0.021 | 0.581 ± 0.072 |

**Table 1.** *Performance of FMCIB+XGBoost and Dinov2+ABMIL Models for KRAS and EGFR Mutation Prediction and TNM Staging on the Stanford Radiogenomics Cohort.* This table summarizes the accuracy, F1-score, sensitivity, specificity, and area under the curve (AUC) for each model, evaluated on KRAS and EGFR mutation status, as well as T and N stages of lung cancer in the Stanford Radiogenomics cohort.

| Stage | Model | Accuracy | F1-Score | Sensitivity | Specificity | AUC |
|---|---|---|---|---|---|---|
| T Stage | FMCIB+XGBoost | 0.726 ± 0.036 | 0.725 ± 0.035 | 0.716 ± 0.052 | 0.735 ± 0.071 | 0.726 ± 0.035 |
| | Dinov2+ABMIL | 0.797 ± 0.038 | 0.795 ± 0.038 | 0.784 ± 0.095 | 0.809 ± 0.087 | **0.797 ± 0.038** |
| N Stage | FMCIB+XGBoost | 0.643 ± 0.053 | 0.641 ± 0.055 | 0.621 ± 0.105 | 0.663 ± 0.040 | 0.642 ± 0.055 |
| | Dinov2+ABMIL | 0.704 ± 0.050 | 0.701 ± 0.052 | 0.664 ± 0.116 | 0.741 ± 0.061 | **0.702 ± 0.052** |

**Table 2.** *Performance of FMCIB+XGBoost and Dinov2+ABMIL Models for TNM Staging on the Lung-CT-PT-Dx Cohort.* This table provides a comparative analysis of accuracy, F1-score, sensitivity, specificity, and AUC for T and N stage predictions in the Lung-CT-PT-Dx cohort using FMCIB+XGBoost and Dinov2+ABMIL models.

## T-Stage Prediction (Stanford Cohort)

For T-stage prediction in the Stanford cohort, FMCIB+XGBoost demonstrated substantially better performance than Dinov2+ABMIL across all evaluated metrics. FMCIB+XGBoost achieved an AUC of 0.855 ± 0.056 and an accuracy of 0.854 ± 0.056, both notably higher than Dinov2+ABMIL's AUC of 0.601 ± 0.055 and accuracy of 0.598 ± 0.061. The F1-score for FMCIB+XGBoost was also significantly higher at 0.853 ± 0.057 compared to 0.588 ± 0.060 for Dinov2+ABMIL, indicating a stronger balance between precision and recall.

FMCIB+XGBoost exhibited well-balanced sensitivity (0.843 ± 0.069) and specificity (0.867 ± 0.067), underscoring its consistent performance in detecting both positive and negative cases of T-stage. In contrast, Dinov2+ABMIL showed lower sensitivity (0.568 ± 0.178) and specificity (0.634 ± 0.148), suggesting less reliable predictions and limited applicability for this staging task.

## N-Stage Prediction (Stanford Cohort)

For N-stage prediction in the Stanford cohort, FMCIB+XGBoost demonstrated consistently better performance compared to Dinov2+ABMIL across all key metrics. FMCIB+XGBoost achieved an AUC of 0.772 ± 0.087 and an accuracy of 0.890 ± 0.042, significantly surpassing Dinov2+ABMIL's AUC of 0.581 ± 0.072 and accuracy of 0.798 ± 0.034. The F1-score for FMCIB+XGBoost was also higher at 0.808 ± 0.087, indicating a more effective balance between precision and recall compared to Dinov2+ABMIL's F1-score of 0.576 ± 0.104.

In terms of sensitivity, FMCIB+XGBoost achieved 0.558 ± 0.173, considerably better than Dinov2+ABMIL's sensitivity of 0.186 ± 0.149. Both models maintained high specificity, with FMCIB+XGBoost slightly outperforming at 0.987 ± 0.016 versus Dinov2+ABMIL's 0.975 ± 0.021. These results highlight the superior ability of FMCIB+XGBoost to detect both positive and negative cases of N-stage, while Dinov2+ABMIL demonstrated limitations in sensitivity and overall performance for this task.

## T-Stage Predictions (Lung-CT-PT-Dx Cohort)

Conversely, in T-stage predictions on the Lung-CT-PT-Dx cohort, Dinov2+ABMIL demonstrated superior performance compared to FMCIB+XGBoost across most metrics. Dinov2+ABMIL achieved an AUC of 0.797 ± 0.038 and an accuracy of 0.797 ± 0.038, outperforming FMCIB+XGBoost, which attained an AUC of 0.726 ± 0.035 and an accuracy of 0.726 ± 0.036. Sensitivity for Dinov2+ABMIL was higher at 0.784 ± 0.095, indicating better detection of positive cases, compared to FMCIB+XGBoost's sensitivity of 0.716 ± 0.052. Similarly, Dinov2+ABMIL demonstrated better specificity (0.809 ± 0.087 vs. 0.735 ± 0.071), showing stronger performance in correctly identifying negative cases. These results highlight Dinov2+ABMIL's generalizability and robustness for T-stage prediction on this diverse dataset.

## N-Stage Predictions (Lung-CT-PT-Dx Cohort)

For N-stage predictions, Dinov2+ABMIL continued to outperform FMCIB+XGBoost. Dinov2+ABMIL achieved an AUC of 0.702 ± 0.052 and an accuracy of 0.704 ± 0.050, exceeding FMCIB+XGBoost's AUC of 0.642 ± 0.055 and accuracy of 0.643 ± 0.053. Dinov2+ABMIL also had higher specificity (0.741 ± 0.061 vs. 0.663 ± 0.040), reflecting its stronger ability to correctly identify negative cases. Sensitivity for Dinov2+ABMIL was slightly improved at 0.664 ± 0.116 compared to FMCIB+XGBoost's 0.621 ± 0.105. These results suggest that Dinov2+ABMIL's SSL-based approach effectively leverages its general training to adapt to this staging task, and it yields more balanced predictions across metrics.

## Discussion

This study presents a comparative analysis of two machine learning models for detecting key lung cancer mutations and staging using 3D nodule data from CT scans. The models evaluated were FMCIB+XGBoost, which utilizes a foundation model with domain-specific pretraining, and Dinov2+ABMIL, which leverages self-supervised learning (SSL) and attention-based multiple-instance learning. Our findings indicate that FMCIB+XGBoost significantly outperformed Dinov2+ABMIL in detecting KRAS and EGFR mutations within the Stanford Radiogenomics cohort, achieving accuracies of 0.846 and 0.883, respectively. In contrast, Dinov2+ABMIL demonstrated superior generalization capabilities in T-stage prediction on the Lung-CT-PT-Dx cohort, achieving an accuracy of 0.797, surpassing FMCIB+XGBoost's accuracy of 0.726. These results underscore the efficacy of domain-specific foundation models in mutation detection tasks and highlight the potential of SSL models in enhancing generalization for cancer staging across diverse datasets.

Prior studies in radiogenomics have often relied on models that require large, annotated datasets for mutation detection and cancer staging tasks. Such models may struggle with generalizability when applied to external datasets due to overfitting and variability in medical imaging data. Our approach builds upon this foundation by incorporating an SSL model, Dinov2+ABMIL, which leverages unlabeled data to learn robust features without the need for extensive annotation. The superior performance of FMCIB+XGBoost in mutation detection aligns with existing literature emphasizing the importance of domain-specific pretraining for specialized tasks. However, the competitive performance of Dinov2+ABMIL in T-stage prediction suggests that SSL models trained on general image datasets can adapt effectively to medical imaging tasks, providing a meaningful advancement over traditional methods.

Accurate detection of KRAS and EGFR mutations is critical for guiding targeted therapies in lung cancer management, traditionally requiring invasive biopsy procedures. Our findings suggest that models like FMCIB+XGBoost, which are pretrained on radiology-specific data, can non-invasively detect these mutations from CT scans with high accuracy, potentially reducing the need for biopsies and associated patient risks. Additionally, the ability of Dinov2+ABMIL to generalize across different datasets for cancer staging tasks indicates its utility in diverse clinical settings, where variations in imaging protocols and patient populations are common. Integrating these models into clinical workflows could enhance personalized treatment planning, improve diagnostic efficiency, and optimize resource allocation in healthcare systems.

A key strength of this study is the use of two independent and diverse cohorts—the Stanford Radiogenomics and Lung-CT-PT-Dx datasets—which enabled a robust assessment of model performance and generalizability. The comparative evaluation of models with domain-specific pretraining and those leveraging self-supervised learning provides valuable insights into their respective advantages in different radiogenomic tasks. By employing 3D nodule data and focusing on both mutation detection and cancer staging, the study offers a comprehensive analysis that mirrors real-world clinical challenges. Furthermore, the use of attention-based multiple-instance learning in the SSL model addresses the common issue of limited or absent instance-level labels in medical imaging data.

Despite the promising results, several limitations warrant consideration. First, the relatively small sample size, particularly for mutation-positive cases, may limit the generalizability of the findings and affect the models' sensitivity. Second, the Dinov2+ABMIL model was pretrained on general natural images, which may not capture domain-specific features essential for medical imaging tasks, potentially impacting its performance in mutation detection. Third, the study's retrospective design means that prospective validation in clinical settings is necessary to confirm the models' utility and reliability. Additionally, the deep learning models used are inherently "black boxes," lacking interpretability, which poses challenges for clinical adoption where understanding the decision-making process is crucial.

In summary, our study demonstrates the complementary strengths of models with domain-specific pretraining and those leveraging self-supervised learning in lung cancer radiogenomics. The FMCIB+XGBoost model's superior performance in mutation detection underscores the value of foundation models trained on radiology data for developing accurate, non-invasive biomarkers. Conversely, the Dinov2+ABMIL model's ability to generalize effectively in cancer staging tasks highlights the potential of SSL approaches to adapt across diverse clinical datasets. These findings contribute to advancing machine learning applications in medical imaging, with promising implications for enhancing personalized medicine and improving patient outcomes in lung cancer care.

# Methods

## Study Cohorts

This study draws from two distinct cohorts to evaluate model performance across mutation detection and cancer staging tasks:

- **Stanford Radiogenomics Cohort:** This cohort comprises annotated 3D lung nodule images, labeled with mutation statuses (KRAS, EGFR) and cancer staging classifications (T-stage, N-stage). Data augmentation strategies, including random rotation and shifting, where the rotation degree was chosen from minus plus 20° and the shifting range is minus plus 15 pixels, were employed to enhance model generalizability across diverse radiogenomic profiles.
- **Lung-CT-PT-Dx Cohort:** Derived from the TCIA database, this cohort consists of 355 3D lung nodule images, with minimal augmentation to retain the original image fidelity. Labeling for cancer staging (T stage and N stage) was confirmed through independent double review by experienced radiologists. Unlike the Stanford cohort, this dataset prioritizes fidelity to clinical imaging settings and serves as a validation set to test model generalizability.
- For both dataset, the same preprocessing steps are taken - standardized voxel resampling (32x32x32) was implemented for model compatibility, with slice thicknesses adjusted to 1.0 mm.

## Model Architectures

- **FMCIB+XGBoost**: In this study, we investigated the effectiveness of foundation models in enhancing deep-learning-based imaging biomarker development, particularly in limited dataset scenarios. The foundation model, FMCIB, is a convolutional encoder pretrained in a self-supervised manner on 11,467 annotated CT lesions from 2,312 unique patients. It was initially validated by classifying lesion anatomical sites, underscoring its robustness in feature extraction.

    For our tasks, we used FMCIB as a feature extractor, without fine-tuning the model itself. The extracted features were subsequently used as input for an XGBoost classifier, fine-tuned for mutation and stage classification. A grid search was conducted to optimize XGBoost's hyperparameters, with the following values: boosting rounds (50, 100, 150), maximum tree depth (3, 5, 7), learning rate (0.01, 0.1, 0.2), and subsampling rate (70%, 80%, 90%). This approach allowed us to leverage FMCIB's foundational representations while performing task-specific fine-tuning within the XGBoost classifier.

- **Dinov2+ABMIL:** Dinov2, a self-supervised learning (SSL) model pre-trained on ImageNet, was further adapted to lung cancer detection using Attention-Based Multiple Instance Learning (ABMIL). ABMIL enables the model to process 3D nodule data without requiring instance-level labels, making it suitable for cases where only aggregate labels are available. Only the ABMIL classifier was trained for mutation and staging classification tasks, while the pretrained weights of Dinov2 remained fixed. This configuration allowed Dinov2 to serve as a feature extractor, with ABMIL leveraging these features to handle cases without instance-level labels effectively.

## Experimental Design

Both models, FMCIB+XGBoost and Dinov2+ABMIL, were evaluated for mutation detection (KRAS, EGFR) and cancer staging (T-stage, N-stage) using a 5-fold cross-validation framework. Stratified sampling within each fold ensured representative distributions of mutation and staging labels, improving generalizability.

For each fold, data from both cohorts was split into training and testing sets at an 80:20 ratio, with cross-validation repeated five times. This design provided separate performance assessments on the Stanford Radiogenomics Cohort and the Lung-CT-PT-Dx Cohort.

Performance was assessed using accuracy, F1-score, sensitivity, specificity, and area under the curve (AUC).